\documentclass[english,aps,prd,
groupedaddress,amssymb,
showpacs,nofootinbib]{revtex4}

\usepackage{times}
\usepackage{amsmath}
\usepackage{graphicx}

\newcommand{\beq}{\begin{eqnarray}}
\newcommand{\eeq}{\end{eqnarray}}

\makeatletter
\numberwithin{equation}{section}
\numberwithin{figure}{section}

\makeatother

\usepackage{babel}

\begin{document}
\setlength{\unitlength}{1mm}

\title{Dynamical Mass Reduction in the Massive Yang-Mills Spectrum in $1+1$ dimensions}

\author{Axel \surname{Cort\'es Cubero}}

\email{acortes_cubero@gc.cuny.edu}

\author{Peter \surname{Orland}}

\email{orland@nbi.dk}

\affiliation{ Baruch College, The 
City University of New York, 17 Lexington Avenue, 
New 
York, NY 10010, U.S.A. }

\affiliation{ The Graduate School and University Center, The City University of New York, 365 Fifth Avenue,
New York, NY 10016, U.S.A.}

\begin{abstract}
The $(1+1)$-dimensional ${\rm SU}(N)$
Yang-Mills Lagrangian, with bare mass $\mathcal M$, and gauge coupling $e$, naively 
describes gluons of mass $\mathcal M$. In fact, renormalization
forces $\mathcal M$ to infinity. The system is in a confined phase, instead of a
Higgs phase. The spectrum of this diverging-bare-mass theory 
contains particles of {\em finite} mass. There are an infinite number of physical particles, which 
are confined hadron-like bound states of 
fundamental colored 
excitations. These particles transform under irreducible representations of the global subgroup of the explicitly-broken
gauge symmetry. The fundamental
excitations are those of the 
${\rm SU}(N)\times {\rm SU}(N)$ principal chiral sigma model, with coupling $g_{0}=e/{\mathcal M}$. We find the  
masses of meson-like bound states of two elementary excitations. This is done using the 
exact S matrix of the sigma model. We point out that the color-singlet spectrum coincides with 
that of the weakly-coupled anisotropic SU($N$)
gauge theory in $2+1$ dimensions. We also briefly comment on how the spectrum behaves in the 't~Hooft limit, $N\rightarrow \infty$.
\end{abstract}

\maketitle

\section{Introduction}

Yang-Mills theory in $1+1$ dimensions has no local degrees of freedom. Introducing
an explicit mass $\mathcal M$ gives a theory of longitudinally-polarized gluons at tree level. It may
seem intuitively obvious, for small gauge coupling, that a particle is either a vector 
Boson, with a mass roughly equal to $\mathcal M$,
or a bound state of such
vector Bosons. This intuition, however, is wrong. We show in this paper that the massive
Yang-Mills theory describes an infinite number of particles, with masses that 
are much less than $\mathcal M$. This can be
called dynamical mass reduction.

Alternatively, the massive Yang-Mills model
can be thought of as a gauge field, coupled to an ${\rm SU}(N)\times {\rm SU}(N)$ principal chiral nonlinear sigma model. The equivalence is seen
by choosing the unitary gauge condition. In a perturbative treatment, the spin waves of the sigma model are Goldstone bosons, giving
the vector particles a mass through the Higgs mechanism. Bardeen and Shizuya used this formulation 
in their proof of renormalizability \cite{bardeen}.	

The tree-level description fails because the excitations of the sigma model
(without the gauge field)
are not Goldstone Bosons. These excitations are massive. Introducing
a gauge field produces a confining force between these excitations. There is no Higgs or Coulomb phase. There is 
only a confined phase.

We briefly describe some important earlier investigations of $(1+1)$-dimensional Yang-Mills theory.  Non-Abelian gauge theories coupled 
to adjoint matter were studied with light-cone
methods by Dalley and Klebanov \cite{DK}. This led to further investigations
of gauged massive adjoint fermions \cite{BHK}. Some detailed results for the spectrum of
the model with
of adjoint scalars were found later \cite{DKB}. Conformal-field-theory methods have recently been applied to the model
with adjoint Fermions \cite{KMTX}.  Much has also been learned about pure Yang-Mills theory in $1+1$ dimensions \cite{Doug}, and
its connections with representation theory.

Our model differs from the Bosonic matter theory of Refs. \cite{BHK}, \cite{DKB},  in
that the matter field has a non-trivial self-interaction. This means that there are two scales in our problem; the mass gap of the sigma model
and the gauge coupling. This is why a nonrelativistic analysis, in which the former is assumed much larger than the latter, can work. A full-fledged relativistic analysis is harder, though we discuss this
problem in the last section of this paper. We wish to stress that we are not studying a massive deformation of pure Yang-Mills theory \cite{Doug} at all. In fact, the situation is exactly the opposite. The deformation is the Yang-Mills action, not the mass term.

A quantum field theory of an SU($N$) gauge field, coupled minimally to an adjoint matter field, can have distinct
Higgs and confinement phases \cite{FradShenk}, separated by a phase boundary, for space-time dimension greater than two. If this
dimension is two, however, there 
is only the confined phase. In the confined phase, the excitations are bound states of the 
massive particles of the 
sigma model. These massive particles are color multiplets of degeneracy $N^{2}$ \cite{wiegmann}.

The action of the massive SU($N$) Yang-Mills field in $1+1$ dimensions is
\beq
S=\int d^2x \left(-\frac{1}{4}{\rm Tr}\,F_{\mu\nu}F^{\mu\nu} +\frac{e^2}{2g_0^2}{\rm Tr} \,A_\mu A^\mu
\right),\label{unitarygauge}
\eeq
where $A_{\mu}$ is Hermitian and 
$F_{\mu\nu}=\partial_\mu A_\nu-\partial_\nu A_\mu -{\rm i}e[A_\mu,A_\nu]$ with $\mu,\nu=0,1$ and
indices are raised by $\eta^{\mu\nu}$, where $\eta^{00}=-\eta^{11}=1,\,\eta^{01}=\eta^{10}=0$. If we drop the cubic and
quartic terms from (\ref{unitarygauge}), the particles are gluons with mass ${\mathcal M}=e/g_{0}$.

Let's now consider a closely-related field theory, namely the ungauged principal chiral sigma model, with action 
\beq
S_{\rm PCSM}=\int d^2x\, \frac{1}{2g_0^2} \,{\rm Tr}\,\partial_\mu U^\dag(x) \partial^\mu U(x),\label{actionpcsm}
\eeq
where the field $U(x)$ is in the fundamental representation of ${\rm SU}(N)$. The action (\ref{actionpcsm}) has a global ${\rm SU}(N)\times {\rm SU}(N)$ symmetry, given by the transformation $U(x)\to V_L U(x) V_R$, where $V_{L,R}\in {\rm SU}(N)$. This model is asymptotically free, and has a mass gap, which we call $m$. It is possible 
that this mass gap is generated by non-real saddle points 
of the functional integral \cite{unsal}. The running bare coupling $g_{0}$ is driven to zero, as the ultraviolet cut-off
is removed.

We promote the left-handed ${\rm SU}(N)$ global symmetry of the sigma model to a local symmetry, by 
introducing the covariant derivative $D_\mu=\partial_\mu-{\rm i} e A_{\mu}$,
where $A_{\mu}$ is a new Hermitian vector field that 
transforms as $A_{\mu}\to V_{L}^{\dag}(x) A_{\mu} V_{L}(x)-\frac{i}{e}V_L^{\dag}(x) \partial_{\mu} V_{L}(x)$. We do not 
gauge the right-handed symmetry. The action is now 
\beq
S=\int d^2x \left[-\frac{1}{4}{\rm Tr} \, F_{\mu\nu} F^{\mu\nu} +\frac{1}{2g_0^2} {\rm Tr}\,(D_\mu U)^{\dag} D^\mu U\right].
\label{gaugedaction}
\eeq
In the unitary gauge, with $U(x)=1$ everywhere, this action (\ref{gaugedaction}) reduces to (\ref{unitarygauge}). In the remainder 
of this paper, however, we will study (\ref{gaugedaction}) 
in the axial gauge.

In our opinion, it is best to think of the left-handed symmetry as (confined) color-SU($N$) and the
right-handed symmetry as flavor-SU($N$). Confinement 
of left-handed color means that only singlets of the left-handed color group exist in the spectrum. There
are ``mesonic" bound states, as well as ``baryonic" bound states. The mesonic bound states have one elementary
particle of the sigma model and one elementary antiparticle. The simplest baryonic bound states consist
of $N$ of these elementary particles, with no antiparticles. There are also more complicated bound states, which
exist because there are excitations in the sigma model (with no gauge
field) 
transforming as higher representations of the color group \cite{wiegmann}. In this paper, we
only discuss the mesonic states in detail.

Recently Gongyo and Zwanziger have studied the nearest-neighbor lattice version of the action (\ref{gaugedaction}) using
Monte-Carlo simulations \cite{GongyoZw}. They computed the
static potential (through the Wilson loop) at different values of the coupling. They find clear evidence of confinement and
string breaking at small values of $g_{0}^{-2}$ (this is proportional to the parameter $\gamma$, in their notation), but a nearly-flat potential
at large values, closer to the continuum limit. They suggest their results may indicate a phase transition to a Higgs phase 
(although they do not assert that this is the case). We believe the explanation is 
the essential singularity of the mass gap as a function of the bare coupling. This mass, in 
an asymptotically-free theory, vanishes faster than any power of of $g_{0}$ as $g_{0}\rightarrow 0$. Thus, string breaking occurs so 
readily, that it may be difficult
to distinguish the two phases. In this paper, the distinction is clear, because we take very small gauge coupling, suppressing (though not eliminating)
string breaking. The continuum gauge
coupling $e$ (with dimensions of mass) is assumed to be much smaller than the mass gap of the sigma model. There should be no phase transition as
the gauge coupling is increased. We therefore
expect that, for any gauge coupling and any value of $g_{0}$, there is only the confined phase. Gongyo and Zwanziger also computed the vector-Boson
propagator (the two-point function of a composite field),
and the order parameter $U$ (in a particular gauge) and the susceptibility of the latter. The lightest bound-state masses could be
found in the behavior of the vector-Boson propagator. This would make for an interesting comparison with our results.

A mesonic bound state, in the
axial gauge, is a sigma-model particle-antiparticle pair, confined by a linear potential. The string tension is
\beq
\sigma={e^2}C_N,\label{stringtension}
\eeq
where $C_N$ is the smallest eigenvalue of the Casimir operator of ${\rm SU}(N)$. The mass gap is 
\beq
M=2m+E_{0}\ll{\mathcal M},\nonumber
\eeq
where $E_{0}$ is the smallest (positive) binding energy, and $m$ is the mass of a sigma-model elementary excitation. This mass $M$ is finite, for
fixed $m$, as the ultraviolet cut-off is removed. In contrast, the bare Yang-Mills mass
$\mathcal M$, which is proportional to $1/g_{0}$, diverges.

Our approach is similar to that of Ref. \cite{orland}. We find the wave function of an unbound particle-antiparticle pair, taking into account scattering
at the origin. Next, we generalize this to the wave function of the pair, confined by a 
linear potential. The method is 
inspired by the determination of the spectrum of the two-dimensional Ising model in 
an external magnetic field \cite{mccoy}. More sophisticated approaches to this and other two-dimensional models of confinement
\cite{DMS}, \cite{DM}, \cite{fonseca}, including fine structure (form factors) of the fundamental excitations, have been developed. We do not take into account decays or corrections to
the spectrum from matrix elements with more fundamental excitations \cite{DGM} in this paper. 
For a general review, see Ref. \cite{tsvelik}. 

We briefly introduce the axial gauge formulation in the next section. In Section III we discuss the S-matrix of the 
principal chiral nonlinear sigma model, and find the free particle-antiparticle wave function, for color group SU($N$), for $N>2$. In Section IV, we find the 
wave functions and bound-state spectrum of a confined pair, for $N>2$ (including $N\rightarrow \infty$ \cite{'t1+1}). We note
that the results generalize the result of Ref. \cite{orland}, on the spectrum of $2+1$-dimensional anisotropic SU($2$) gauge theories, to SU($N$). We treat the $N=2$ case separately in Section V. We present some conclusions 
and proposals for further work in the last section.

\section{The axial gauge formulation and the confined phase}

Care is necessary to understand why the bare mass is not the physical mass. If the axial
gauge $A_{1}=0$, is chosen, the action (\ref{gaugedaction}) is
\beq
S=\int d^{2}x \left[\frac{1}{2}{\rm Tr}\,(\partial_{1}A_{0})^{2}+
\frac{1}{2g_{0}^{2}}{\rm Tr}\,(\partial_{0}U^{\dagger}+{\rm i}eU^{\dagger}A_{0})(\partial_{0}U-{\rm i}eA_{0}U)
-\frac{1}{2g_{0}^{2}}{\rm Tr}\,\partial_{1}U^{\dagger}\partial_{1}U
 \right]\,. \nonumber
\eeq
Let us introduce the traceless Hermitian generators $t_{a}$ of SU($N$), $a=1,\dots, N^{2}-1$, with normalization 
${\rm Tr}\,t_{a}t_{b}=\delta_{ab}$ and structure coefficients $f_{abc}$, defined by $[t_{b},t_{c}]={\rm i}f_{abc}t_{a}$. If we 
naively eliminate $A_{0}$, by its equation of motion (or integrate $A_{0}$ from the functional integral), we obtain the effective
action
\beq
S=\int d^{2}x \left(\frac{1}{2g_0^2} {\rm Tr}\,\partial_\mu U^{\dag} \partial^\mu U +\frac{1}{2} \,{j_{0}^{L}}_{a}\,\frac{1}{-\partial_{1}^{2}+e^{2}/g_{0}^{2}}
\, {j_{0}^{L}}_{a}\right), \label{screened}
\eeq
where $j_\mu^{L}(x)_{b}=-{\rm i} {\rm Tr}\,t_{b} \partial_\mu U(x) U^\dag(x)$
is the Noether current of the left-handed ${\rm SU}(N)$ symmetry. The potential induced on the color-charge density,
in the second term of (\ref{screened}), indicates that charges are screened, instead of confined. This conclusion, however, 
is based on the fact that $U^{\dagger}U=1$. In
the renormalized theory, $U$ is not a physical field. The physical scaling field of the principal chiral nonlinear sigma model is not
a unitary matrix. This fact is discussed more explicitly in Refs. \cite{tHPCM}, in the limit $N\rightarrow\infty$, with $g_{0}^{2}N$ 
fixed. The actual excitations of the principal
chiral model are massive, with a left and right color charge \cite{wiegmann}, so that no screening takes place.

A more careful approach is to first find the Hamiltonian in the temporal gauge $A_{0}=0$. Gauge invariance, or Gauss' law, must be imposed on
physical states. The Hamiltonian is
\beq
H=\int dx^{1} \,\left\{\frac{g_{0}^{2}}{2} [j^{L}_{0}(x^{1})_{b}]^{2}+ \frac{1}{2g_{0}^{2}}[j^{L}_{1}(x^{1})_{b}]^{2}+ \frac{1}{2}[E(x^{1})_{b}]^{2}
+\frac{e}{g_{0}^{2}} j^{L}_{1}(x^{1})_{b} A_{1}(x^{1})_{b}\right\},    \label{axialgauge}
\eeq
where $A_{1}(x^{1})_{b}={\rm Tr}\,t_{b}A$ and $E_{a}$ is the electric field, obeying $[E(x^{1})_{a}, A_{1}(y^{1})_{b}]=-{\rm i}\delta_{ab}\delta(x^{1}-y^{1})$. The Hamiltonian
(\ref{axialgauge}) must be 
supplemented by Gauss' law $G(x^{1})_{a}\Psi=0$, for any physical state $\Psi$, where $G(x^{1})_{a}$ is the generator of spatial gauge transformations:
\beq
G(x^{1})_{a}=\partial_{1}E(x^{1})_{a}+e f_{abc}A_{1}(x^{1})^{b}E(x^{1})_{c}-\frac{e}{g_{0}^{2}} j^{L}_{0}(x^{1})_{a}\,.
\eeq
If we require that the electric field vanishes at the boundaries $x^{1}=\pm l/2$, Gauss' law may be explicitly solved \cite{2+1}, to yield the expression for the electric field:
\beq
E(x^{1})_{a}=\int_{-l/2}^{x^{1}}dy^{1} \,\left\{ {\mathcal P}\exp\left[ ie\int_{-l/2}^{y^{1}} dz^{1}{\mathcal A}_{1}(z^{1})\right]\right\}_{a}^{\;\;\;\;\;b}\;\;
\frac{e}{g_{0}^{2}} j^{L}_{0}(y^{1})_{b}, \label{electricequiv}
\eeq
where ${\mathcal A}_{1}(x^{1})_{a}^{\;\;\;b}={\rm i}f_{abc}A_{1}(x^{1})_{c}$ is the gauge field in the adjoint representation. There remains a global gauge invariance, which must be satisfied by physical states, {\em i.e.}, $\Gamma_{a}\Psi=0$, where
\beq
\Gamma_{a}= \int_{-l/2}^{l/2}dy^{1} \,\left\{ {\mathcal P}\exp\left[ ie\int_{-l/2}^{y^{1}} dz^{1}{\mathcal A}_{1}(z^{1})\right]\right\}_{a}^{\;\;\;\;\;b}\;\;
\frac{e}{g_{0}^{2}} j^{L}_{0}(y^{1})_{b} .
\label{residual}
\eeq
Now we are free to chose $A_{1}(x^{1})_{b}=0$, which simplifies (\ref{electricequiv}) and (\ref{residual}). The solution for the
electric field yields the Hamiltonian
\beq
H=\int dx^{1} \,\left\{\frac{g_{0}^{2}}{2} [j^{L}_{0}(x^{1})_{b}]^{2}+ \frac{1}{2g_{0}^{2}}[j^{L}_{1}(x^{1})_{b}]^{2}\right\}
-\frac{e^{2}}{2g_{0}^{4}} \int dx^{1} \!\!\int dy^{1}\; \vert x^{1}-y^{1} \vert \; j^{L}_{0}(x^{1})_{b} \; j^{L}_{0}(y^{1})_{b}
,    \label{finalaxialgauge}
\eeq
where in the last step, we have taken the size $l$ of the system to infinity. The last term is a linear potential which confines left-handed color. Notice
that (\ref{finalaxialgauge}) is not bounded from below on the full Hilbert space. This is because of the last, nonlocal term; the energy can be lowered by
adding pairs of colored particles (or antiparticles) and by separating them. The residual Gauss-law
condition $\Gamma_{a}\Psi=0$, forces the global left-handed color to be a singlet, thereby removing the instability,

\section{The Free Particle-Antiparticle Wave Function: $N>2$}

The quantized principal chiral nonlinear sigma model is integrable. This property, together with physical considerations, has been 
used to find the exact S-matrix \cite{wiegmann}.

An excitation has rapidity $\theta$, related to that excitation's energy and momentum, by $E=m\sinh\theta$ and $p=m\cosh\theta$, respectively.

Let us consider a state with two excitations. One excitation is an antiparticle of rapidity $\theta_1$ and left and right ${\rm SU}(N)$ color indices $a_1,b_1=1,\dots, N$, respectively. The second excitation is a particle of rapidity $\theta_2$, and left and right color indices $a_2,b_2$, respectively. Explicitly
the state is
\beq
\vert A, \theta_1, b_1,a_1; P, \theta_2,a_2,b_2\rangle_{\rm in}.\nonumber
\eeq
The S-matrix element, $S(\theta)_{a_1b_1;b_2a_2}^{d_2c_2;c_1d_1}$, is defined by
\beq
\,_{\rm out}\langle A, \theta^\prime_1,d_1,c_1; P,\theta^\prime_2,c_2,d_2\vert A, \theta_1, b_1,a_1; P, \theta_2,a_2,b_2\rangle_{\rm in}=S(\theta)_{a_1b_1;b_2a_2}^{d_2c_2;c_1d_1}\,4\pi \delta(\theta_1-\theta^\prime_1)\,4\pi \delta(\theta_2-\theta^\prime_2),\nonumber
\eeq
where $\theta=\theta_1-\theta_2$. This S-matrix element is \cite{wiegmann}
\beq
S(\theta)_{a_1b_1;b_2a_2}^{d_2c_2;c_1d_1}=S(\theta)\left[\delta_{a_1}^{c_1}\delta_{a_2}^{c_2}-\frac{2\pi {\rm i}}{N(\pi {\rm i}-\theta)}\delta_{a_1a_2}\delta^{c_1c_2}\right]\left[\delta_{b_1}^{d_1}\delta_{b_2}^{d_2}-\frac{2\pi {\rm i}}{N(\pi {\rm i}-\theta)}\delta_{b_1b_2}b^{d_1d_2}\right],\nonumber
\eeq
where
\beq
S(\theta)=\frac{\sinh\left[\frac{(\pi {\rm i}-\theta)}{2}-\frac{\pi {\rm i}}{N}\right]}{\sinh\left[\frac{(\pi {\rm i}-\theta)}{2}+\frac{\pi {\rm i}}{N}\right]}\,\left\{\frac{\Gamma[i(\pi {\rm i}-\theta)/2\pi+1]\Gamma[-{\rm i}(\pi {\rm i}-\theta)/2\pi-{1}/{N}]}{\Gamma[{\rm i}(\pi {\rm i}-\theta)/2\pi+1-1/N]\Gamma[-{\rm i}(\pi {\rm i}-\theta)/2\pi]}\right\}^2.\label{wiegmann}
\eeq
For $N>2$, the expression (\ref{wiegmann}) may be written in the exponential form \cite{cubero} :
\beq
S(\theta)=\exp2 \int_0^\infty\,\frac{d\xi}{\xi \sinh\xi}\left[ 2(e^{2\xi/N}-1)-\sinh(2\xi/N)
\right]\sinh\frac{\xi \theta}{\pi{\rm i}} \;.
\label{Stheta}
\eeq
We will discuss the $N=2$ case separately in Section V.

The wave function of a free antiparticle at $x^1$ and a free particle at $x^{2}$, with momenta $p_1$ and $p_2$, respectively, is 
\beq
\Psi_{p_1,\,p_2}(x^1,y^1)_{a_1a_2;b_1b_2}=\left\{\begin{array}{cc}
e^{{\rm i}p_1x^1+{\rm i}p_2y^1}A_{a_1a_2;b_1b_2},\;&{\rm for} \;x^1<y^1,\\ \\
e^{{\rm i}p_2x^1+{\rm i}p_1y^1}S(\theta)_{a_1b_1;b_2a_2}^{d_2c_2;c_1d_1}A_{c_1c_2;d_1d_2},\; &{\rm for}\; x^1>y^1.\end{array}
\right. \label{freewavefunction}
\eeq
where $A_{a_1a_2;b_1b_2}$ is set of arbitrary complex numbers.

The residual Gauss' law in the axial gauge, $\Gamma_{a}\Psi=0$, restricts physical states to those which are invariant under global
left-handed ${\rm SU}(N)$ color transformations.  This means that the 
particle-antiparticle state of the form (\ref{freewavefunction}) must be projected to a global left-color singlet. A left-color-singlet 
wave function is
\beq
\!\!\!\!\!\!\!\!\!\!\!\!
\Psi_{p_1p_2} (x^1,y^1)_{b_1b_2}=\delta^{a_1a_2}\Psi_{p_1,\,p_2}(x^1,y^1)_{a_1a_2b_1b_2}.
\label{leftsinglet}
\eeq

There are states of degeneracy $N^{2}-1$, which resemble massive gluons. These transform as the adjoint representation
of the right-handed color symmetry. The wave function of such a state 
is traceless in the right-handed color indices:
\beq
\delta^{b_1b_2}\Psi_{p_1p_2}\!\!\!&(&\!\!\!x^1,y^1)_{b_1b_2}=0.\label{tracelesscondition}
\eeq
We use a non-relativistic approximation $p_{1,2}\ll m$. The wave function in this limit becomes
\beq
\Psi_{p_1p_2}(x^1,y^1)_{b_1b_2}=\left\{\begin{array}{c}
e^{{\rm i}p_1x^1+{\rm i}p_2y^1}A_{b_1b_2},\,\,\,\,\,\,\,\,\,\,\,\,\,{\rm for}\,\,x^1<y^1,\\
\,\\
e^{{\rm i}p_2x^1+{\rm i}p_1y^1}\exp ({\rm i}\pi -\frac{i h_{N}}{\pi m}\vert p_1-p_2\vert )A_{b_1b_2},\,\,\,\,\,\,{\rm for}\,\,x^1>y^1.\end{array}\right.
\label{nonrelativisticwave}
\eeq
where ${\rm Tr} A=0$, and
\beq
h_{N}&=&2\int_0^\infty \frac{d\xi}{\sinh \xi}\left[2(e^{2\xi/N}-1)-\sinh(2\xi/N)\right]\nonumber\\
&=&-4\gamma-\psi\left(\frac{1}{2}+\frac{1}{N}\right)-3\psi\left(\frac{1}{2}-\frac{1}{N}\right)-4\ln 4,
\label{N>2}
\eeq
where $\gamma$ is the Euler-Mascheroni constant, and 
$\psi(x)={d}\ln \Gamma(x)/dx$ is the digamma function. The 
expression in (\ref{nonrelativisticwave}) must be equal to the wave function of two confined particles for
sufficiently small $\vert x^1-y^1\vert$. To compare the two expressions, it is convenient to use center-of-mass coordinates, $X,\,x$, and their respective momenta $P,\,p$. Explicitly, 
$X=x^{1}+y^{1}$, $x=y^{1}-x^{1}$, $P=p_{1}+p_{2}$ and $p=p_{2}-p_{1}$. In these coordinates, the wave function is
\beq
\Psi_p(x)_{b_1b_2}=\left\{\begin{array}{c}
\cos(px+\omega)A_{b_1b_2},\,\,\,\,\,\,\,\,\,\,\,{\rm for}\,\,x>0,\\
\,\\
\cos[-px+\omega-\phi(p)]A_{b_1b_2},\,\,\,\,\,\,\,{\rm for}\,\,x<0,\end{array}\right.\label{comcoordinates}
\eeq
for some constant $\omega$, with the phase shift $\phi(p)=\pi-\frac{h_N}{\pi m}\vert p\vert$.

Another type of mesonic state is the right-handed color singlet, with $A_{b_1b_2}=\delta_{b_1b_2}$. The non-relativistic limit 
of the wave function in this case is
\beq
\Psi_p(x)_{\rm singlet}=\left\{\begin{array}{c}
\cos(px+\omega),\,\,\,\,\,\,\,\,\,\,\,{\rm for}\,\,x>0,\\
\,\\
\cos[-px+\omega-\chi(p)],\,\,\,\,\,\,\,{\rm for}\,\,x<0,\end{array}\right.\label{singletstate}
\eeq
where $\chi(p)=-\frac{h_N}{\pi m}\vert p\vert.$

\section{Mesonic States of Massive Yang-Mills Theory: $N>2$}

The wave function of a particle-antiparticle pair, confined by string tension $\sigma$, satisfies the Schroedinger equation
\beq
-\frac{1}{m}\frac{d^2}{dx^2}\Psi(x)_{b_1b_2}+\sigma \left\vert x \right\vert \,\Psi(x)_{b_1b_2}=E\Psi(x)_{b_1b_2},\label{schroedinger}
\eeq
where $E$ is the binding energy \cite{mccoy}. The solution to Equation (\ref{schroedinger}) is
\beq
\Psi(x)_{b_1b_2}=\left\{\begin{array}{c}
C {\rm Ai}\left[(m\sigma)^{\frac{1}{3}}\left(x+\frac{E}{\sigma}\right)\right]A_{b_1b_2},\,\,\,\,\,\,\,\,\,{\rm for}\,\,x>0\\
\,\\
C^\prime {\rm Ai}\left[(m\sigma)^{\frac{1}{3}}\left(-x+\frac{E}{\sigma}\right)\right]A_{b_1b_2},\,\,\,\,\,\,\,\,{\rm for}\,\,x<0,\end{array}\right.\label{airy}
\eeq
where ${\rm Ai}(x)$ is the Airy function of the first kind, and $C,\,C^\prime$ are constants.

For $\vert x\vert\ll (m\sigma)^{-1/3}$, the potential energy in (\ref{schroedinger}) is sufficiently small that the wave function is 
(\ref{comcoordinates}), with $\vert p\vert=(mE)^{\frac{1}{2}}$. The wave function (\ref{airy}) is approximated in this region by
\beq
\Psi(x)_{b_1b_2}=\left\{\begin{array}{c}
C\frac{1}{\left(x+\frac{E}{\sigma}\right)^{\frac{1}{4}}}\cos \left[\frac{2}{3}(m\sigma)^{\frac{1}{2}}\left(x+\frac{E}{\sigma}\right)^{\frac{3}{2}}-\frac{\pi}{4}\right] A_{b_1b_2},\,\,\,\,\,\,\,{\rm for}\,\,x>0,\\
\,\\
C^\prime\frac{1}{\left(-x+\frac{E}{\sigma}\right)^{\frac{1}{4}}}\cos \left[-\frac{2}{3}(m\sigma)^{\frac{1}{2}}
\left(-x+\frac{E}{\sigma}\right)^{\frac{3}{2}}+\frac{\pi}{4}\right]A_{b_1b_2},\,\,\,\,\,\,\,{\rm for}\,\,x<0.\end{array}\right.\nonumber
\eeq

Let us now consider the $(N^{2}-1)$-plet of mesonic states.  The wave functions (\ref{comcoordinates}) and (\ref{airy}) should be the same for $x\downarrow0$, yielding
\beq
\frac{C}{(\frac{E}{\sigma})^{\frac{1}{4}}}\cos\left[\frac{2}{3}(m\sigma)^{\frac{1}{2}}\left(\frac{E}{\sigma}\right)^{\frac{3}{2}}-\frac{\pi}{4}\right]=\cos(\omega).\label{conditionabove}
\eeq
Equation (\ref{conditionabove}) implies
\beq
C=\left(\frac{E}{\sigma}\right)^{\frac{1}{4}},\,\,\,\,\,\,\,\omega=\frac{2}{3}(m\sigma)^{\frac{1}{2}}\left(\frac{E}{\sigma}\right)^{\frac{3}{2}}-\frac{\pi}{4}.\nonumber
\eeq
The wave functions (\ref{comcoordinates}) and (\ref{airy}) should also be the same for $x\uparrow0$, yielding
\beq
\frac{C^\prime}{\left(\frac{E}{\sigma}\right)^{\frac{1}{4}}}\cos\left[-\frac{2}{3}(m\sigma)^{\frac{1}{2}}\left(\frac{E}{\sigma}\right)^{\frac{3}{2}}+\frac{\pi}{4}\right]=\cos \left[\omega-\pi+\frac{h_N}{\pi m}(mE)^{\frac{1}{2}}\right],\label{fromabove}
\eeq
hence $C^\prime=C=\left(\frac{E}{\sigma}\right)^{\frac{1}{4}}$. The arguments of the cosine on each side of (\ref{fromabove}) must be the same, modulo $2\pi$:
\beq
-\frac{2}{3}(m\sigma)^{\frac{1}{2}}\left(\frac{E}{\sigma}\right)^{\frac{3}{2}}+\frac{\pi}{4}+2\pi n=\frac{2}{3}(m\sigma)^{\frac{1}{2}}\left(\frac{E}{\sigma}\right)^{\frac{3}{2}}-\frac{5\pi}{4}+\frac{h_N}{\pi m}(m E)^{\frac{1}{2}},\nonumber
\eeq
for $n=0,1,2,\dots$. We simplify this to
\beq
\frac{4}{3}(m\sigma)^{\frac{1}{2}}\left(\frac{E}{\sigma}\right)^{\frac{3}{2}}+\frac{h_N}{\pi m}(m E)^{\frac{1}{2}}-\left(n+\frac{3}{4}\right)2\pi=0.\label{quantization}
\eeq

An analysis which is similar to that of the previous paragraph yields the quantization condition for the right-handed singlet state (\ref{singletstate}). This is 
\beq
\frac{4}{3}(m\sigma)^{\frac{1}{2}}\left(\frac{E}{\sigma}\right)^{\frac{3}{2}}+\frac{h_N}{\pi m}(m E)^{\frac{1}{2}}-\left(n+\frac{1}{4}\right)2\pi=0.\label{quantizationsinglet}
\eeq

Equations (\ref{quantization}) and (\ref{quantizationsinglet}) are depressed cubic equations of the variable $Z_n=E_n^{\frac{1}{2}}$. These cubic equations have only one real solution for each value of $n$, because ${h_{N}}/({\pi m^{\frac{1}{2}}})>0$. The solution of Equations (\ref{quantization}) and (\ref{quantizationsinglet}) is
\beq
E_n=\left\{\left[\epsilon_n+\left(\epsilon_n^2+\beta_N^3\right)^{\frac{1}{2}}\right]^{\frac{1}{3}}+\left[\epsilon_n-\left(\epsilon_n^2+\beta_N^3\right)^{\frac{1}{2}}\right]^{\frac{1}{3}}\right\}^{\frac{1}{2}},\label{spectrum1} \eeq
where
\beq
\epsilon_n=\frac{3\pi}{4}\left(\frac{\sigma}{m}\right)^{\frac{1}{2}}\left(n+\frac{1}{2}\pm \frac{1}{4}\right),\,\,\,\,\,\,\,\,\,\,\,\,\,
\beta_N=\frac{h_N\sigma^{\frac{1}{2}}}{4\pi m}, \label{spectrum2}
\eeq
where $\pm=+$ for the $(N^2-1)$-plet, and $\pm=-$ for the singlet.

We show in the next section that the expressions (\ref{spectrum1}) and (\ref{spectrum2}) remain valid for the SU($2$) case, with 
$h_{2}=-4\ln 2+2$ and, significantly, with a reversal of the sign in (\ref{spectrum2}). For $N=2$ {\em only} we must take
$\pm=-$ for the $(N^{2}-1)$-plet (the triplet) and $\pm=+$ for the singlet. 

As it happens, the results we have just obtained for the singlet spectrum  
generalize the result of Ref. \cite{orland}, on the spectrum of $2+1$-dimensional anisotropic SU($2$) gauge theories, to SU($N$) (where $\sigma$ is
replaced by $2\sigma$). 

Another interesting special case is the 't~Hooft limit $N\rightarrow \infty$ \cite{tHPCM}, \cite{correlations}. The mass gap of the sigma model should be fixed in this
limit. The string tension $\sigma$ will be fixed as well \cite{'t1+1}, provided $e^{2}N$ is fixed.  In this limit $h_{N}\rightarrow 0$, and we find
\beq
E_n=\left[\frac{3\pi}{2}\left(\frac{\sigma}{m}\right)^{\frac{1}{2}}\left(n+\frac{1}{2}\pm \frac{1}{4}\right) \right]^{1/3}.
\label{spectrumlargeN}
\eeq

\section{The $N=2$ case}

The exponential expression for the S-matrix (\ref{Stheta}) is only correct for $N>2$. The principal chiral model with $SU(2)\times SU(2)$ symmetry is equivalent to the $O(4)$-symmetric nonlinear sigma model. We will express the S matrix, first found in Ref. \cite{zamolodchikov}, by an exponential expression \cite{karowski}.

A state with one excitation has a left-handed color index $a=1,2$ and a right-handed color index $b=1,2$. In the $O(4)$ formulation, excitations have a single species index $j=1,2,3,4$. The $SU(2)\times SU(2)$-symmetric states are related to the $O(4)$-symmetric states by
\beq
\vert P, \theta, a, b\rangle_{\rm in}&=&\sum_j \frac{1}{\sqrt{2}} \left(\delta^{j 4}\delta_{ab}-i\sigma_{ab}^j\right) \vert \theta, j\rangle_{\rm in},\nonumber\\
\vert A, \theta, a, b\rangle_{\rm in}&=&\sum_j \frac{1}{\sqrt{2}} \left(\delta^{j 4}\delta_{ab}-i\sigma_{ab}^j\right)^* \vert \theta, j\rangle_{\rm in},\nonumber
\eeq
where $\sigma^j$ with $j=1,2,3$ are the Pauli matrices. The $O(4)$ two-excitation S-matrix,  $S(\theta)^{j_1 j_2}_{j^\prime_1 j^\prime_2} $ is given by
\beq
\,_{\rm out}\langle \theta^\prime_1, j^\prime_1;\theta^\prime_2, j^\prime_2\vert \theta_1, j_1;\theta_2,j_2\rangle_{\rm in}\nonumber=S(\theta)^{j_1 j_2}_{j^\prime_1 j^\prime_2}\, 4\pi \delta(\theta_1-\theta^\prime_1)\,4\pi \delta(\theta_2-\theta^\prime_2),\nonumber
\eeq
where \cite{karowski}
\beq
S(\theta)^{j_1 j_2}_{j^\prime_1 j^\prime_2} = \left[\frac{\theta+\pi i}{\theta-\pi i}(P^0)^{j_1 j_2}_{j_1^\prime j_2^\prime}+\frac{\theta-\pi i}{\theta+\pi i} (P^{+})^{j_1 j_2}_{j_1^\prime j_2^\prime}+(P^{-})^{j_1 j_2}_{j_1^\prime j_2^\prime}\right] Q(\theta),\nonumber
\eeq
\beq
Q(\theta)=\exp 2\int_0^\infty \frac{d\xi}{\xi}\frac{e^{-\xi}-1}{e^\xi+1}\sinh\left(\frac{\xi\theta}{\pi {\rm i}}\right),\nonumber
\eeq
and $P^0,\,P^{+},$ and $P^{-}$ are the singlet, symmetric-traceless, and antisymmetric projectors, which are
\beq
(P^0)^{j_1 j_2}_{j_1^\prime j_2^\prime}=\frac{1}{4}\delta^{j_1j_2}\delta_{j_1^\prime j_2^\prime}&,&\,\,\,
(P^+)^{j_1 j_2}_{j_1^\prime j_2^\prime}=\frac{1}{2}(\delta^{j_1}_{j_1^\prime}\delta^{j_2}_{j_2^\prime}+\delta^{j_1}_{j_2^\prime}\delta^{j_2}_{j_1^\prime})
-\frac{1}{4}\delta^{j_1j_2}\delta_{j_1^\prime j_2^\prime},\nonumber\\
(P^-)^{j_1 j_2}_{j_1^\prime j_2^\prime}&=&\frac{1}{2}(\delta^{j_1}_{j_1^\prime}\delta^{j_2}_{j_2^\prime}-\delta^{j_1}_{j_2^\prime}\delta^{j_2}_{j_1^\prime}),\nonumber
\eeq
respectively.

We write the left-color-singlet wave function for a free particle and antiparticle:
\beq
\Psi_{p_1,p_2}(x^1,y^1)_{b_1b_2}&=&D_{b_1b_2}^{j_1j_2}\left\{\begin{array}{c}
e^{ip_1 x^1+ip_2y^1}A_{j_1 j_2},\,\,\,\,{\rm for}\,\,x^1>y^1\\
\,\\
e^{ip_2x^1+ip_1y^1}S(\theta)^{j^\prime_1 j^\prime_2}_{j_1 j_2} A_{j^\prime_1 j^\prime_2},\,\,\,\,{\rm for}\,\,x^1<y^1,
\end{array}\right.\label{ofourwave}
\eeq
where
\beq
D_{b_1b_2}^{j_1j_2}=\frac{1}{2}\delta^{a_1a_2}\left(\delta^{j_1 4}\delta_{a_1b_1}-i\sigma_{a_1b_1}^{j_1}\right)^*\left(\delta^{j_2 4}\delta_{a_2b_2}-i\sigma_{a_2b_2}^{j_2}\right)\,.\nonumber
\eeq
There is a triplet of degenerate states and one singlet state. The triplet satisfies
\beq
\delta^{b_1b_2}\Psi_{p_1,p_2}(x^1,y^1)_{b_1b_2}=0.\label{tracelessofour}
\eeq
Substituting (\ref{ofourwave}) into (\ref{tracelessofour}) gives the condition
\beq
\delta^{b_1b_2}\,D_{b_1b_2}^{j_1j_2}\,A_{j_1j_2}=\delta^{j_1 j_2}A_{j_1j_2}=0\,.\nonumber
\eeq
The traceless matrix $A_{j_1j_2}$ can be split into a symmetric and an antisymmetric part, $A^{+}_{j_1j_2}=(A_{j_1j_2}+A_{j_2j_1})/2$ and $A^{-}_{j_1j_2}=(A_{j_1j_2}-A_{j_2j_1})/2$, respectively. The matrix $A^{+}_{j_1j_2}$, however, does not contribute to the wave function (\ref{ofourwave}), because
\beq
D^{j_1j_2}_{b_1b_2}A^{+}_{j_1j_2}=\frac{1}{2}\delta_{b_1b_2}{\rm Tr}\,A^{+}=0.\nonumber
\eeq
The matrix $A^{-}_{j_1 j_2}$ satisfies \cite{zamolodchikov}, \cite{karowski}:
\beq
S(\theta)^{j_{1} j_{2}}_{j_{1}^{\prime} j_{2}^{\prime}}A^{-}_{j_{1}j_{2}}=Q(\theta)A^{-}_{j_{1}^{\prime}j_{2}^{\prime}}.\label{antisymmetric}
\eeq
Substituting (\ref{antisymmetric}) into (\ref{ofourwave}), in center-of-mass coordinates and the non-relativistic limit, we find
\beq
\Psi_{p}(x)_{b_1b_2}=D_{b_1b_2}^{j_1j_2}\left\{\begin{array}{c}
\cos(px+\omega)A_{j_1j_2},\,\,\,\,\,\,\,\,\,\,\,{\rm for}\,\,x>0,\\
\,\\
\cos[-px+\omega-\phi(p)]A_{j_1j_2},\,\,\,\,\,\,\,{\rm for}\,\,x<0,\end{array}\right.\label{nonrelativisticofour}
\eeq
where 
$\phi(p)=-\frac{i h_2}{\pi m}\vert p\vert$, where
\beq
h_2=2\int_0^\infty d\xi\,\frac{e^{-\xi}-1}{e^\xi+1}=-4\ln 2+2.\label{N=2}
\eeq

The wave function of the right-color-singlet bound state is
\beq
\Psi_{p_1,p_2}^{\rm singlet}(x^1,y^1)=
\left\{\begin{array}{c}
e^{ip_1 x^1+ip_2y^1}
,\,\,\,\,{\rm for}\,\,x^1>y^1,\\
\,\\
e^{ip_2x^1+ip_1y^1}\frac{\theta+\pi {\rm i}}{\theta-\pi {\rm i}}Q(\theta)
,\,\,\,\,{\rm for}\,\,x^1<y^1.
\end{array}\right.\label{singletofourwave}
\eeq
In center-of-mass coordinates, in the non-relativistic approximation, this becomes
\beq
\Psi_{p}^{\rm singlet}(x)=
\left\{\begin{array}{c}
\cos(px+\omega),\,\,\,\,\,\,\,\,\,\,\,{\rm for}\,\,x>0,\\
\,\\
\cos[-px+\omega-\chi(p)],\,\,\,\,\,\,\,{\rm for}\,\,x<0,\end{array}\right.\label{nonrelativisticofoursinglet}
\eeq
where $\chi(p)=\pi -\frac{i h_2}{\pi m}\vert p\vert$.

From this point onward, the analysis is similar to what we've presented in the last two sections. We obtain (\ref{spectrum1}), (\ref{spectrum2}),
except that $h_{N}$ (defined in (\ref{N>2})) is replaced with $h_{2}$ (defined in (\ref{N=2})), with one important difference; we
have $\pm=+$ for the singlet and $\pm=-$ for the triplet in Eq. (\ref{spectrum2}). As mentioned at the end of the last section, the
singlet spectrum coincides with that of Ref. \cite{orland}, in which $\sigma$ must be replaced by $2\sigma$.

\section{Conclusions and Outlook}

We have found the spectrum of massive $(1+1)$-dimensional SU($N$) Yang-Mills theory, for small gauge coupling. To do this, we formulated
the model as
a principal chiral sigma model coupled to a massless Yang-Mills field. In the axial gauge, there are sigma-model particles and antiparticles which
bind to make left-color singlets. We obtained the mesonic spectrum by determining the particle-antiparticle wave function in the non-relativistic limit, taking into account the phase shift at the origin.

In the future, we would like to find relativistic corrections to the mass spectrum. This was done in Ref. \cite{fonseca} for the Ising model in 
an external magnetic field. The goal would be to find mesonic eigenstates of the Hamiltonian (\ref{finalaxialgauge}) of the form:
\beq
\vert \Psi_B\rangle_{b_1b_2}=\vert \Psi_B^{(2)}\rangle_{b_1b_2}+\vert \Psi_B^{(4)}\rangle_{b_1b_2}+\vert \Psi_B^{(6)}\rangle_{b_1b_2}+\dots,\nonumber
\eeq
where the state $\vert \Psi_B^{(2M)}\rangle_{b_1b_2}$ contains $M$ particles and $M$ antiparticles. The multi-particle contributions are included
because an electric string may break \cite{DGM}, producing pairs of sigma-model excitations. Nonetheless, for
small gauge coupling, the ``two-quark" approximation is valid. In the this approximation, the bound state is treated as
\beq
\vert \Psi_B\rangle_{b_1b_2}\approx\vert \Psi_B^{(2)}\rangle_{b_1b_2}&=&\frac{1}{2}\int
\frac{d\theta_{1}}{4\pi}
\frac{d\theta_{2}}{4\pi}
\Psi(p_1,p_2)_{a_2a_2}\vert A,\theta_1,b_1,a_1;P,\theta_2,a_2,b_2\rangle,\;
{\rm where},\nonumber \\ 
\Psi(p_1,p_2)_{a_1a_2}&=&S(\theta)\left[\delta_{a_1}^{c_1}\delta_{a_2}^{c_2}-\frac{2\pi {\rm i}}{N(\pi {\rm i}-\theta)}\delta_{a_1a_2}\delta^{c_1c_2}\right]\Psi(p_2,p_1)_{c_1c_2}.\label{twoquark}\
\eeq
The spectrum of masses $\Delta$, of the states (\ref{twoquark}) is found from 
the Bethe-Salpeter equation $(H-\Delta)\vert \Psi^{(2)}_B\rangle_{b_1b_2}=0$. Acting on this state with the Hamiltonian (\ref{finalaxialgauge}) yields
\beq
\left(m\cosh\theta_1+m\cosh\theta_2-\Delta\right)\!\!\!\!\!&&\!\!\!\!\!\Psi(p_1^\prime,p_2^\prime)_{c_1c_2}\delta_{b_1d_1}\delta_{b_2d_2}\nonumber\\
&=&\frac{e^2}{4g_0^4}\int
\frac{d\theta_{1}}{4\pi}
\frac{d\theta_{2}}{4\pi}
\Psi(p_1,p_2)_{a_1a_2}\int dx^1 dy^1\vert x^1-y^1\vert\nonumber\\
&&\times\langle A, \theta_1^\prime, d_1,c_1;P,\theta_2^\prime,c_2,d_2\vert {\rm Tr}\,\left[j_0^L(x^1)j_0^L(y^1)\right]\vert A,\theta_1, b_1, a_1;P, \theta_2,a_2,b_2\rangle,\label{bethesalpeter}
\eeq
where the operator ${\rm Tr}\,\left[j_0^L(x^1)j_0^L(y^1)\right]$ is not time-ordered. The matrix element
\beq
\langle A, \theta_1^\prime, d_1,c_1;P,\theta_2^\prime,c_2,d_2\vert {\rm Tr}\,\left[j_0^L(x^1)j_0^L(y^1)\right] \vert A,\theta_1, b_1, a_1;P, \theta_2,a_2,b_2\rangle\nonumber
\eeq
is obtained by inserting a complete set of states between the current operators and using the exact 
form factors of the currents of the principal chiral sigma model. For finite $N$, only the leading two-particle form factors of currents are known \cite{cubero}
and only a vacuum insertion can be made. The complete matrix element is known at large $N$ \cite{correlations}, which should
help in finding the relativistic corrections to the eigenvalues of Eq. (\ref{bethesalpeter}).

\begin{acknowledgements}
A.C.C. would like to thank Davide Gaiotto and Jaume Gomis for interesting discussions, and the Perimeter Institute for their hospitality. P.O.'s work was supported by a grant from the PSC-CUNY.
\end{acknowledgements}

\end{document}